\documentclass{jkas}

\def\year{2016} 
\def\month{August} 
\def\volume{49} 
\def\issue{4} 
\def\beginpage{101} 
\def\endpage{112} 
\def\received{June 30, 2016} 
\def\accepted{August **, 2016} 
\date{Received \received; accepted \accepted}

\def\kms{~{\rm km~s^{-1}}}
\def\cm3{~{\rm cm^{-3}}}

\def\muG{~{\mu\rm G}}

\usepackage{flushend}

\title{Re-acceleration model for the `Sausage' Radio Relic}

\author{Hyesung Kang}

\affil{Department of Earth Sciences, Pusan National University, Pusan 609-735, Korea; \email{hskang@pusan.ac.kr}}

\def\corrauthor{H. Kang}

\def\runningauthor{Kang}

\def\runningtitle{Re-Acceleration Model for Radio Relics}

\def\keywords{acceleration of particles --- cosmic rays ---
galaxies: clusters: general --- shock waves}

\def\abstracttext{
The Sausage radio relic is the arc-like radio structure in the cluster CIZA
J2242.8+5301, whose observed properties 
can be best understood by synchrotron emission from relativistic electrons 
accelerated at a merger-driven shock. 
However, there remain a few puzzles that cannot be explained by the shock acceleration model with
only in-situ injection. In particular, the Mach number inferred from the observed radio spectral index, 
$M_{\rm radio}\approx 4.6$, while the Mach number estimated from X-ray observations, $M_{\rm X-ray}\approx 2.7$.
In an attempt to resolve such a discrepancy, here we consider the re-acceleration model in which a shock of 
$M_s\approx 3$ sweeps through the intracluster gas with a pre-existing population of relativistic electrons.
We find that observed brightness profiles at multi frequencies provide strong constraints on the spectral shape of
pre-existing electrons.
The models with a power-law momentum spectrum with the slope, $s\approx 4.1$, and the cutoff Lorentz
factor, $\gamma_{e,c}\approx 3-5\times 10^4$ can reproduce reasonably well the observed spatial profiles of radio fluxes 
and integrated radio spectrum of the Sausage relic.
The possible origins of such relativistic electrons in the intracluster medium remain to be investigated further.
}

\begin{document}
\jkashead 

\section{Introduction}

Giant radio relics such as the Sausage and the Toothbrush relics
exhibit elongated morphologies, spectral steepening across the relic width,
integrated radio spectra of a power-law form with spectral curvature above $\sim 2 $~GHz, 
and high polarization level \citep{vanweeren10,vanweeren12, feretti12, stroe16}.
They are thought to be synchrotron radiation emitted by GeV electrons, which are (re)-accelerated
at structure formation shocks in the intracluster medium (ICM)
\citep[e.g.,][]{ensslin98,  brug12, brunetti2014}. 
It is now well established that nonthermal particles can be (re)-accelerated 
at such shocks via diffusive shock acceleration (DSA)  process 
\citep[e.g.,][]{ryu03,vazza09,skill11,kangryu11}.

In the simple DSA model of a steady planar shock,
the synchrotron radiation spectrum at the shock becomes a power-law of
$j_{\nu}(r_s)\propto \nu^{-\alpha_{\rm sh}}$ with the {\it shock index}, $\alpha_{\rm sh} = (M_s^2+3)/2(M_s^2-1)$,
while the volume-integrated radio spectrum becomes $J_{\nu} \propto \nu^{-\alpha_{\rm int}}$ 
with the integrated index, $\alpha_{\rm int}=\alpha_{\rm sh}+0.5$, above the break frequency $\nu_{\rm br}$
\citep[e.g.][]{dru83,ensslin98,kang11}.
Here $M_s$ is the shock sonic Mach number.
If the shock acceleration duration is less than $\sim 100$ Myr, however, 
the break frequency, $\nu_{\rm br}\sim 1$ GHz, falls in the typical observation frequencies and
the integrated spectrum steepens gradually over the frequency range of $(0.1-10) \nu_{\rm br}$ \citep{kang15b}. 
Moreover, additional spectral curvatures can be introduced in the case of a spherically expanding shock \citep{kang15a}.
On the other hand,
in the re-acceleration model in which the upstream gas contains a pre-existing electron population, for example,
$f_{\rm up} \propto \gamma_e^{-s} \exp [ - ({\gamma_e /\gamma_{e,c}})^2 ]$ (where $\gamma_e$ is the Lorentz factor),
the re-accelerated electron spectrum and the ensuing radio spectrum should depend on the slope $s$ and the cutoff energy $\gamma_{e,c}$ as well as $M_s$.

Recently, \citet{kang16} (Paper I) has explored the observed properties of the Tooth brush relic, 
and also reviewed some puzzles in the DSA origin of radio {\it gischt} relics:
(1) discrepancy between $M_{\rm radio}$ inferred from the radio index $\alpha_{\rm sh}$ and $M_{\rm X-ray}$ estimated from the
X-ray temperature discontinuities in some relics,
(2) low DSA efficiency expected for weak shocks with $M_s\lesssim 3$ that form in the hot ICM,
(3) low frequency of merging clusters with detected radio relics, compared to the expected occurrence of
ICM shocks,
and (4) shocks detected in X-ray observations without associated radio emission.
In Paper I, it was suggested that most of these puzzles can be explained by the re-acceleration model in which a radio
relic lights up only when a shock propagates in the ICM thermal plasma that contains a pre-existing population of electrons \citep[see also][]{kang12,pinzke13, shimwell15,kangryu16}.

The so-called Sausage relic is a giant radio relic in the outskirts of the merging cluster CIZA
J2242.8+5301, first detected by \citet{vanweeren10}.
They interpreted the observed radio spectrum from 150 ~MHz to 2.3 GHz as a power-law-like synchrotron 
radiation emitted by shock-accelerated relativistic electrons.
So they inferred the shock Mach number, $M_{\rm radio}\approx 4.6$, from the spectral index at the hypothesized shock location, 
$\alpha_{\rm sh}\approx 0.6$,
and the magnetic field strength, $B_2\approx 5$ or $1.2 \muG$ from the relic width of 55~kpc.
Although this shock interpretation was strongly supported by observed downstream spectral aging and high polarization levels,
the requirement for a relatively high Mach number of $M_s=4.6$ in the ICM called for some concerns.
Based on structure formation simulations, the shocks in the ICM are expected to have low Mach numbers,
typically $M_s<3$ \citep[e.g.,][]{ryu03}.

\citet{stroe14b} reported that the integrated spectrum of the Sausage relic steepens toward 16~GHz with
the integrated index increasing from $\alpha_{\rm int}\approx 1.06$ to $\alpha_{\rm int}\approx 1.33$ above 2.3~GHz.
They noted that such a curved integrated spectrum cannot be consistent with the simple DSA model for a 
steady plane shock with $M_s\approx4.6$ suggested by \citet{vanweeren10}.
Later, \citet{stroe14a} suggested, using a spatially resolved spectral fitting, that the
the injection index could be larger, i.e.,~$\alpha_{\rm sh}\approx 0.77$, implying $M_s\approx 2.9$.
In fact, this lower value of $M_s$ is more consistent with temperature discontinuities detected in
X-ray observations by \citet{ogrean14} and \citet{akamatsu15}.

In order to understand the spectral curvature
in the integrated spectrum reported by \citet{stroe14b},
\citet{kangryu15} considered the various shock models including both the {\it in-situ injection} model 
without pre-existing electrons and
the {\it re-acceleration} model with pre-existing electrons of a power-law spectrum with exponential cutoff.
It was shown that shock models with $M_s\approx 3$, either the {\it in-situ injection} or the {\it re-acceleration} models
can reproduce reasonably well the radio brightness profile at 600~MHz and the curved integrated spectrum of the Sausage relic 
except the abrupt increase of the spectral index above 2~GHz.
The authors concluded that such a steep increase of the spectral index cannot be explained by the simple radiative 
cooling of postshock electrons.
On the other hand, it was pointed out that the Sunyaev-Zeldovich effect may induce such 
spectral steepening by reducing the the radio flux
at high frequencies by a factor of two or so \citep{basu15}.

Recently, \citet{stroe16} presented the integrated spectrum spanning from 150~MHz up to 30~GHz of the Sausage relic,
which exhibits a spectral steepening from $\alpha_{\rm int}\approx 0.9$ at low frequencies to $\alpha_{\rm int}\approx 1.71$ above 2.5~GHz. 
\citet{kangryu16} attempted to reproduce this observed spectrum
by the re-acceleration model in which a spherical shock sweeps through and then
exits out of a finite-size region with pre-existing relativistic electrons.
Since the re-acceleration stops after the shock crosses the region with pre-existing electrons,
the ensuing integrated radio spectrum steepens much more than predicted for aging postshock electrons alone,
resulting in a better match to the observed spectrum. 
We suggested that a shock of $M_s \approx 2.7-3.0$ and $u_s \approx 2.5-2.8\times 10^3 \kms$
that has swept up the cloud of $\sim 130$~kpc with pre-existing electrons about 10 Myr ago, 
could reproduce the observed radio flux profile at 600~MHz \citep{vanweeren10} 
and the observed integrated spectrum \citep{stroe16}.
The required spectral shape of pre-existing electrons is a power-law spectrum with the slope, $s=4.2$
and the exponential cutoff at $\gamma_{e,c}\approx 10^4$.

On the other hand, \citet{donnert16} has proposed an alternative approach to explain the spectral steepening of
the Sausage relic. In order to match the observed brightness profiles,
it was assumed that behind the shock the magnetic field strength increases first, peaks around 40~kpc from the shock
and then decreases exponentially at larger distances.
In this model, the magnetic field strength is lower at the immediate postshock region with the highest energy electrons,
compared to the model with constant postshock magnetic field.
As a result, the integrated radio spectrum steepens at high frequencies, leading to the curved spectrum consistent with
the observation by \citet{stroe16}.

That paper presented beam convolved brightness profiles, $S_{\nu}(R)$, at several radio frequencies 
from 153~MHz to~30 GHz in their Figure 5 and the spectral index, $\alpha_{153}^{608}$ between 153 and 608~MHz in Figure 6.
We notice that $S_{\nu}$ at $153-323$~MHz extends well beyond 150~kpc away from the relic edge,
and $\alpha_{153}^{608}$ increases from $\sim 0.6$ at the position of the
putative shock to $\sim 1.9$ at 200~kpc south of the shock.
Considering the shock compression ratio of $\sigma\approx 3$, these downstream length scales imply 
that the shock has swept through a region of at least $450$~kpc in the case of a plane shock.
So these observations cannot be explained by the model of \citet{kangryu16} which assumed 
a cloud with pre-existing electrons of 130~kpc in width.

As pointed out in Paper I, the ubiquitous presence of radio galaxies, AGN relics and radio phoenix implies that 
the ICM  may contain radio-quiet {\it fossil} electrons ($\gamma_{e,c}\lesssim 10^2$)
or radio-loud {\it live} electrons ($\gamma_{e,c}\lesssim 10^4$) \citep[e.g.,][]{slee01}.
In the re-acceleration model,
fossil electrons with $\gamma_e\sim 100$ provide seed electrons that can be injected to the DSA process 
and enhance the acceleration efficiency at weak ICM shocks.
On the other hand, radio-loud electrons of a power-law spectrum with a cutoff 
at $\gamma_{e,c}\sim 7-8\times 10^4$ is required to explain the broad relic width of $\sim 150-200$~kpc
in the case of the Toothbrush relic (Paper I).

In our re-acceleration model of the Sausage relic, considered in \citet{kangryu15, kangryu16}, 
the shock propagates into the thermal ICM gas
with pre-existing relativistic electrons whose pressure is dynamically insignificant.
Note that here the preshock medium is not a bubble of hot buoyant relativistic plasma
unlike the models studied previously by \citet{ensslin01}, \citet{ensslin02}, and \citet{pfrommer11}.
Thus the presence of pre-existing electrons does not affect the dynamics of the shock, but instead it 
only provides the seed electrons that can be injected effectively into the DSA process.
However, it is not clear how relativistic electrons can be mixed with the thermal ICM gas,
if they were to originate from radio jets and lobes ejected from AGNs.
On the other hand, such a mixture of thermal gas and relativistic electrons can be understood
more naturally, 
if they were to be produced by previous episodes of shocks and turbulence generated 
by merger-driven activities in the ICM \citep[e.g.,][]{brunetti2014}.

In this study, we attempt to explain the observed properties of the Sausage relic,
reported by \citet{stroe16} and \citet{donnert16},
with the re-acceleration model in which a low Mach number shock  
sweeps though the ICM gas with pre-existing relativistic electrons, 
as we did for the Toothbrush relic in Paper I.
In the next section, we explain some basic physics of the DSA model and review
the observed properties of the Sausage relic. 
In Section 3 the numerical simulations and the shock models are described.
The comparison of our results with observations is presented in Section 4,
followed by a brief summary in Section 5.

\begin{figure}[t!]
\centering
\includegraphics[trim=1mm 4mm 4mm 8mm, clip, width=84mm]{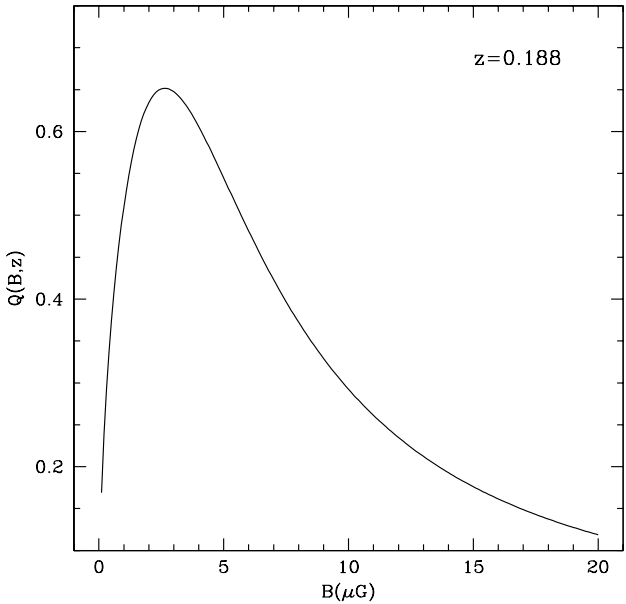}
\caption{The factor $Q(B,z)$ for $z=0.188$ given in Equation~(\ref{qfactor}).
}
\end{figure}

\begin{table*}
\begin{center}
{\bf Table 1.}~~Model Parameters for the Sausage Relic Shock\\
\vskip 0.3cm
\begin{tabular}{ lrrrrrrrrrrr }
\hline\hline

Model & $M_{\rm s,i}$ & $kT_1$ & $B_1$ & $s$ & $\gamma_{\rm e,c}$ & $t_{\rm exit}$ & $L_{\rm cloud}$& $t_{\rm obs}$& $M_{\rm s,obs}$ & $kT_{\rm 2,obs}$ & $u_{\rm s,best}$ \\
 {}  & &{(keV)} &($\muG$) & &   & (Myr)& (kpc) & (Myr)& &{(keV)} & {(${\rm km~s^{-1}}$)}   \\

\hline
M3.3  & 3.3 &3.4& 1& 4.1 & $5\times10^4$ & 124 & 367 & 144 & 2.7  & 10.7 & $2.6\times10^3$  \\
M3.8  & 3.8 &3.4& 1& 4.1 & $3\times10^4$ & 125 & 419 &143 & 3.1  & 13.1 & $2.9\times10^3$ \\

\hline
\end{tabular}
\end{center}
{$M_{\rm s,i}$: initial shock Mach number at the onset of the simulations}\\
{$kT_1$: preshock temperature}\\
{$B_1$: preshock magnetic field strength}\\
{$s$: power-law slope in Equation (2)}\\
{$\gamma_{e,c}$: exponential cutoff in Equation (2)}\\
{$t_{\rm exit}$: time when the shock exists the cloud with pre-existing electrons}\\
{$L_{\rm cloud}$: size of the cloud with pre-existing electrons}\\
{$t_{\rm obs}$: shock age when the simulated results match the observations}\\
{$M_{\rm s,obs}$: shock Mach number at $t_{\rm obs}$}\\
{$kT_{\rm 2,obs}$: postshock temperature at $t_{\rm obs}$}\\
{$u_{\rm s,obs}$: shock speed at $t_{\rm obs}$}\\

\end{table*}

\section{Model}

\subsection{Physics of DSA}

The basic physical features of the DSA model for radio relics were described in detail in Paper I.
Some of them are repeated here in order to make this paper sufficiently self-contained.

In the limit where the electrons that are injected {\it in situ} at a shock with $M_s$
dominate over the re-accelerated electrons, 
the radio synchrotron spectrum at the shock position can be approximated by a power-law of
$j_{\nu}(r_s)\propto \nu^{-\alpha_{\rm sh}}$ with the `shock' index 
\begin{equation}
\alpha_{\rm sh}= {{(M_s^2+3)}\over {2(M_s^2-1)}}.
\label{alpha}
\end{equation}

In the other limit where the re-accelerated electrons dominate over the injected electrons,
the radio synchrotron spectrum is not a simple power-law, but depends not only on the shock Mach number
but also on the spectral shape of pre-existing electrons, for instance,
\begin{equation}
f_{\rm up}(\gamma_e) \propto \gamma_e^{-s} \exp \left[ - \left({\gamma_e \over \gamma_{e,c}} \right)^2 \right].
\label{fexp}
\end{equation}
Then the radio spectral index at low frequencies ($\sim 100$~MHz) emitted by re-accelerated electrons
can be approximated by
$\alpha_{\rm sh}=(s-3)/2$, independent of $M_s$.

Synchrotron and inverse Compton (iC) energy losses introduce gradual spectral steepening in
the volume-integrated radio spectrum behind the shock,
leading to the increase of the `integrated' index, $\alpha_{\rm int}$ from $\alpha_{\rm sh}$ to $\alpha_{\rm sh}+0.5$
over $\sim(0.1-10)\nu_{\rm br}$.
Here the break frequency depend on the magnetic field strength and the shock age as follows:
\begin{equation}
\nu_{\rm br}\approx 0.63 {\rm GHz} \left( {t_{\rm age} \over {100 \rm Myr}} \right)^{-2}
 Q(B,z)^2.
\label{fbr}
\end{equation}
The factor $Q$ is defined as
\begin{equation}
Q(B,z)\equiv \left[ { {(5\muG)^2} \over {B^2+B_{\rm rad}(z)^2}}\right] \left({B \over 5 \muG}\right)^{1/2},
\label{qfactor}
\end{equation}
where $B_{\rm rad}=3.24\muG(1+z)^2$ and $B$ is expressed in units of $\muG$
\citep{kang11}.
Figure 1 shows that $Q$ evaluated for $z=0.188$ has the maximum value, $Q=0.65$ at $B\approx 2.5\muG$.

The profiles of observed radio flux at multi frequencies can provide strong constraints on the model parameters for radio relics.
For instance, the width of radio relics observed at low frequencies, whose radiation comes mainly from
uncooled low energy electrons, should be similar to the advection length: 
\begin{equation}
\Delta l_{\rm low}\approx 
100~{\rm kpc} \cdot W_l \cdot u_{2,3}\cdot \left({{t_{\rm age}}\over {100~{\rm Myr}}}\right),
\label{lwidth1}
\end{equation}
where $u_{2,3}=u_2/10^3 \kms$ and $t_{\rm age}$ is the duration of the shock acceleration.
The factor $W_l\sim 1.1-1.2$ reflects the fact that the downstream flow speed in the shock rest frame
increases  behind the shock in the case of a spherically expanding shock (see Figure 2 below).
For planar shocks, $W_l\approx 1$.

On the other hand, the relic width at high frequencies due to cooled electrons 
becomes similar to the cooling length:
\begin{eqnarray}
\Delta l_{\rm high}
\approx 100\ {\rm kpc} \cdot W_h \cdot u_{2,3}\cdot Q\cdot
\left[{{\nu_{\rm obs}(1+z) \over {0.63{\rm GHz}}} }\right]^{-1/2},
\label{lwidth2}
\end{eqnarray}
where $\nu_{\rm obs}$ is the observation frequency and the factor $W_h\sim 1.2-1.3$ 
takes account for both the downstream speed effects described above and 
the synchrotron radiation emitted by low energy electrons.
Note that the synchrotron emission decreases smoothly behind the shock, 
so Equations (5) and (6) give only characteristic length scales.

\subsection{Injection-dominated versus Re-acceleration dominated Model}

The intracluster space may contain fossil relativistic electrons accelerated by the structure formation shocks,
since the ICM is expected to go through such shocks twice or so on average \citep{ryu03}.
In addition, there could be numerous radio galaxies and 
radio ghosts in the ICM \citep{slee01}.
As a result, some cluster shocks could encounter a cloud with pre-existing relativistic electrons.
Thus one can construct two limiting scenarios to explain the observations of the Sausage relic:
the relic is produced (1) by the shock with $M_s\approx 4.6$ without pre-existing electrons (in-situ injection model),
or (2) by the shock with $M_s\approx 3$ with pre-existing electrons with the power-law index,
$s\approx 4.2$ (re-acceleration model).

In the case where in-situ injection dominates over re-acceleration, 
the shock Mach number should be $M_s\approx 4.6$ to explain the spectral
index, $\alpha_{\rm sh}\approx 0.6 $ at the edge of the Sausage relic.
Then the postshock temperature would be $kT_2= 20.2$~keV (for $kT_1= 2.7$~keV), 
which is well above the observed postshock temperature, $8.5_{-0.6}^{+0.8}$~keV \citep{akamatsu15}.
So we disfavor this in-situ injection model mainly because of the discrepancy in
the postshock temperature.
Moreover, $M_s=4.6$ is rather high for typical shocks ($M_s\lesssim 2$)
that are expected to form in the hot ICM during the course of cluster mergers
\citep[e.g.][]{ryu03}.
However, we note that \citet{donnert16} argue that, based on the in-situ injection model, 
the observed brightness profiles can be modeled only with the postshock flow speed, $u_2\gtrsim 1200\kms$.
For the $M_4=4.6$ model the shock speed becomes $u_s=4.1\times 10^3 \kms$, 
and the downstream flow speed is estimated to be $u_2=1.2\times 10^3 \kms$.

In the re-acceleration model, on the other hand,
one can adjust the pre-existing electron spectrum in order to
reproduce the observed radio brightness profiles and the integrated spectrum.
For example, the power-law index should be $s\approx 4.2$ to match the observed value, 
$\alpha_{\rm sh}\approx 0.6$ at the edge of the Sausage Relic.
One of the advantages of this model is the fact that, independent of $\alpha_{\rm sh}$, 
we can adopt a low Mach number ($M_s\lesssim 3$), 
which is more compatible with X-ray observations.
The cutoff energy, $\gamma_{e,c}$, also controls how fast the electron spectrum steepens behind the shock,
so it can be adjusted to reproduce the observed spectral aging in the downstream region.

\subsection{Observed Properties of the Sausage Relic}

The cluster CIZA J2242.8+5301 that hosts the Sausage relic is estimated to be located at the
redshift, $z=0.188$ \citep{dawson15}.
According to the Suzaku X-ray observations, the ICM temperature drops from
$kT_2= 8.5_{-0.6}^{+0.8}$~keV to $kT_1= 2.7_{-0.4}^{+0.7}$~keV across the relic
with the inferred shock Mach number, $M_s=2.7_{-0.4}^{+0.7}$ \citep{akamatsu15}.
If we take the mean observed values, 
the sound speed of the preshock gas with 2.7~keV is $u_1=8.4\times 10^2\kms$,
so the shock speed is $u_s=2.3\times 10^3 \kms$ for a $M_s=2.7$ shock.
With the compression ratio of $\sigma= u_1/u_2 = 2.83$,
the downstream flow speed becomes $u_2 = 8.0\times 10^2 \kms$,
which is probably too slow to explain the observed relic widths \citep{donnert16}.

The radio observations of the Sausage relic from 150~MHz to 30~GHz using various radio telescopes
were reported by \citet{stroe16}, where the flux data for the integrated spectrum was given in their Table 3.
The brightness profiles of the relic at several radio frequencies based on the observations 
in \citet{stroe16} have been published recently in \citet{donnert16}.
The FWHMs of the observed relic at 153~MHz and 608~MHz, measured from Figure 5 in \citet{donnert16},
are 100~kpc and 55~kpc, respectively.
These radio data can be used to inferred the shock parameters such $M_s$, $u_s$ and $B_2$
according to Equations (1)-(6).
According to Equation (6) with $W_h=1.2$, $u_{2,3}=0.8$, and $Q=0.65$ ($B_2=2.5\muG$), for example,
the relic width at 608~MHz is estimated to be $\Delta l_{\rm high}\approx 58$~kpc,  
which seems to be consistent with the observed profile.
On the other hand, the observed brightness profile at 153~MHz extends up to 200~kpc,
while the predicted cooling length is only $\Delta l_{\rm high}\approx 116$~kpc.
Thus the shock model parameters can be determined more accurately through the detail comparisons 
between the predicted and observed brightness profiles.

According to Figure 1 of \citet{stroe13}, there are at least four radio galaxies, sources B, C, D, and H,
within 1~Mpc from the Sausage relic, which might provide relativistic electrons to the surrounding ICM.
In particular, the source H located at the eastern edge of the relic might be feeding relativistic electrons
to the upstream region of the relic shock. 
It is not clear, however, how those electrons can be mixed with the thermal background gas
as we conjecture in our model.

\section{Numerical Calculations}

The numerical setup for our DSA simulations was described in detail in \citet{kang15b}.
So only basic features are given here.

\subsection{DSA Simulations for 1D Spherical Shocks}

We follow time-dependent diffusion-convection equation
for the pitch-angle-averaged phase space distribution function
for CR electrons, $f_e(r,p,t)=g_e(r,p,t)p^{-4}$, in the one-dimensional (1D) 
spherically symmetric geometry:

\begin{eqnarray}
{\partial g_e\over \partial t}  + u {\partial g_e \over \partial r}
= {1\over{3r^2}} {{\partial (r^2 u) }\over \partial r} \left( {\partial g_e\over
\partial y} -4g_e \right)  \nonumber\\
+ {1 \over r^2}{\partial \over \partial r} \left[ r^2 D(r,p)
{\partial g_e \over \partial r} \right]
+ p {\partial \over {\partial y}} \left( {b\over p^2} g_e \right),
\label{diffcon}
\end{eqnarray}
where $u(r,t)$ is the flow velocity, $y=\ln(p/m_e c)$, $m_e$ is the electron mass, and $c$ is
the speed of light \citep{skill75}.
Here $r$ is the radial distance from the center of the spherical coordinate,
which assumed to coincide with the cluster center.
We assume a Bohm-like spatial diffusion coefficient, $D(r,p)\propto p/B$. 
The cooling term $b(p)=-dp/dt= - p/t_{\rm rad} $ accounts for electron synchrotron and iC losses.
The test-particle version of CRASH (Cosmic-Ray Amr SHock) code in a comoving 1D spherical grid is used to
solve Equation (\ref{diffcon}) \citep{kj06}.

\subsection{Shock parameters}

We assume that the shock dynamics can be approximated by a self-similar blast wave 
that propagate through the isothermal ICM with the density profile of $\rho\propto r^{-2}$.
So the shock radius and velocity evolves roughly as $r_s\propto t^{2/3}$ and $u_s\propto t^{-1/3}$,
respectively \citep[e.g.,][]{ryu91}.

\citet{donnert16} chose the following shock parameters to explain the observed profiles of 
$S_{\nu}$ and $\alpha_{153}^{608}$: $kT_1=3.0$~keV, $M_s=4.6$, and $u_s=4.1\times10^3\kms$. 
Then $\sigma=\rho_2/\rho_1=3.5$, $kT_2=22.4$~keV, and $u_2=1.2\times10^3\kms$.
Although the downstream temperature is well above the observed values, $kT_2=9.57_{-1.12}^{+1.25}$~keV
\citep{ogrean14} or $kT_2=8.5_{-0.6}^{+0.8}$~keV
\citep{akamatsu15}, they adopted the high value of $M_s=4.6$, because it is consistent with
$\alpha_{\rm sh}=0.6$ in the in-situ injection model,
and because the observed flux profiles require the downstream flow speed as large as $1.2\times10^3\kms$.

Here we adopt a different set of shock parameters that may be more consistent with X-ray observations.
The preshock temperature is chosen as $kT_1=3.4$~keV ($c_{s,1}=0.94\times 10^3 \kms$).
Then we choose two values of the initial Mach number (see Table 1):

{\bf M3.3 model}: $M_{s,i}=3.3$, $u_{2,i}=0.99\times 10^3 \kms$, and $\gamma_{e,c}=5\times10^4$. 
 
{\bf M3.8 model}: $M_{s,i}=3.8$, $u_{2,i}=1.1\times 10^3 \kms$, and $\gamma_{e,c}=3\times10^4$.

\begin{figure*}[t]
\centering
\includegraphics[trim=5mm 2mm 5mm 10mm, clip, width=140mm]{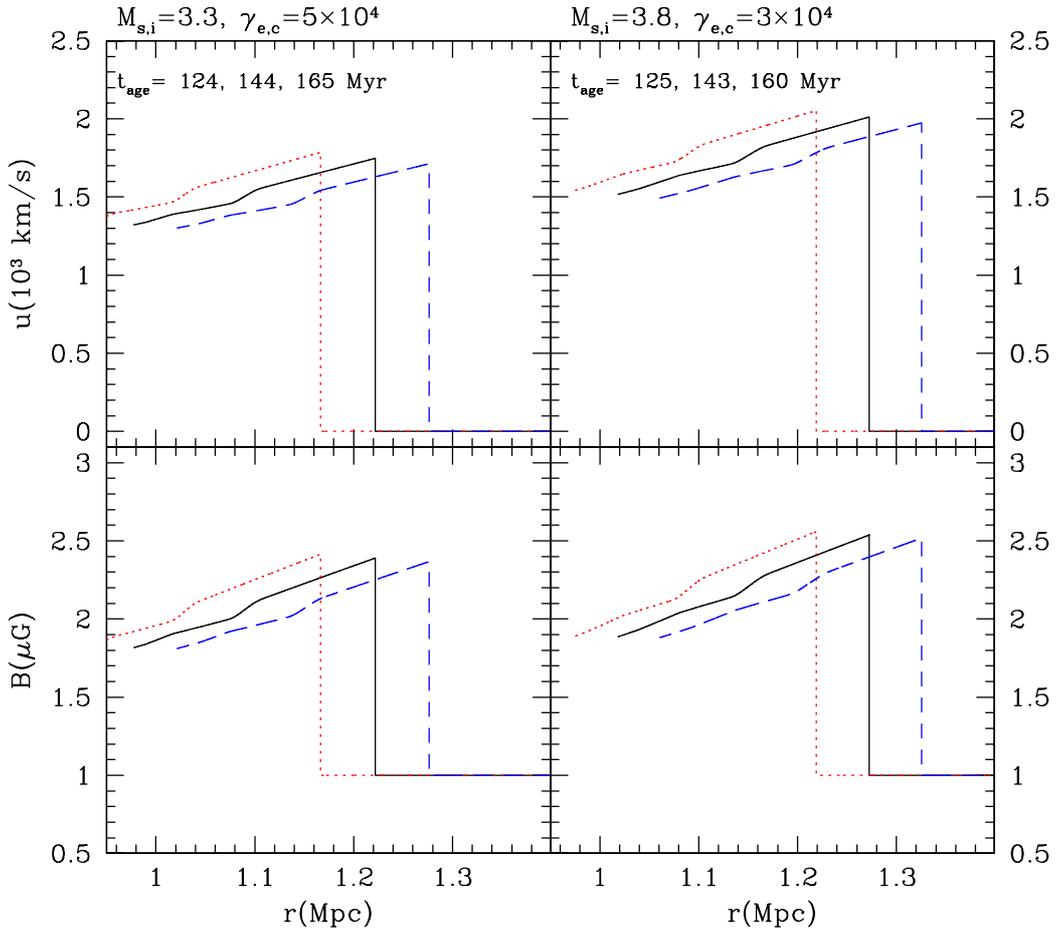}
\vskip -0.5cm
\caption{Flow velocity, $u(r)$, in units of $10^3\kms$ (top panels), and the magnetic field strength, B(r), 
in units of microgauss, plotted as a function of the radial distance from the cluster center, $r(\rm Mpc)$,
for the M3.3 model at $t_{\rm age}=$ 124 (red dotted line), 144 (black solid), and 165 (blue dashed)~Myr (left panels) 
and for the M3.8 model at $t_{\rm age}=$ 125 (red dotted line), 143 (black solid), and 160 (blue dashed)~Myr  (right panels).
}
\end{figure*}
\begin{figure*}[t!]
\vskip -0.5cm
\centering
\includegraphics[trim=2mm 2mm 2mm 2mm, clip, width=150mm]{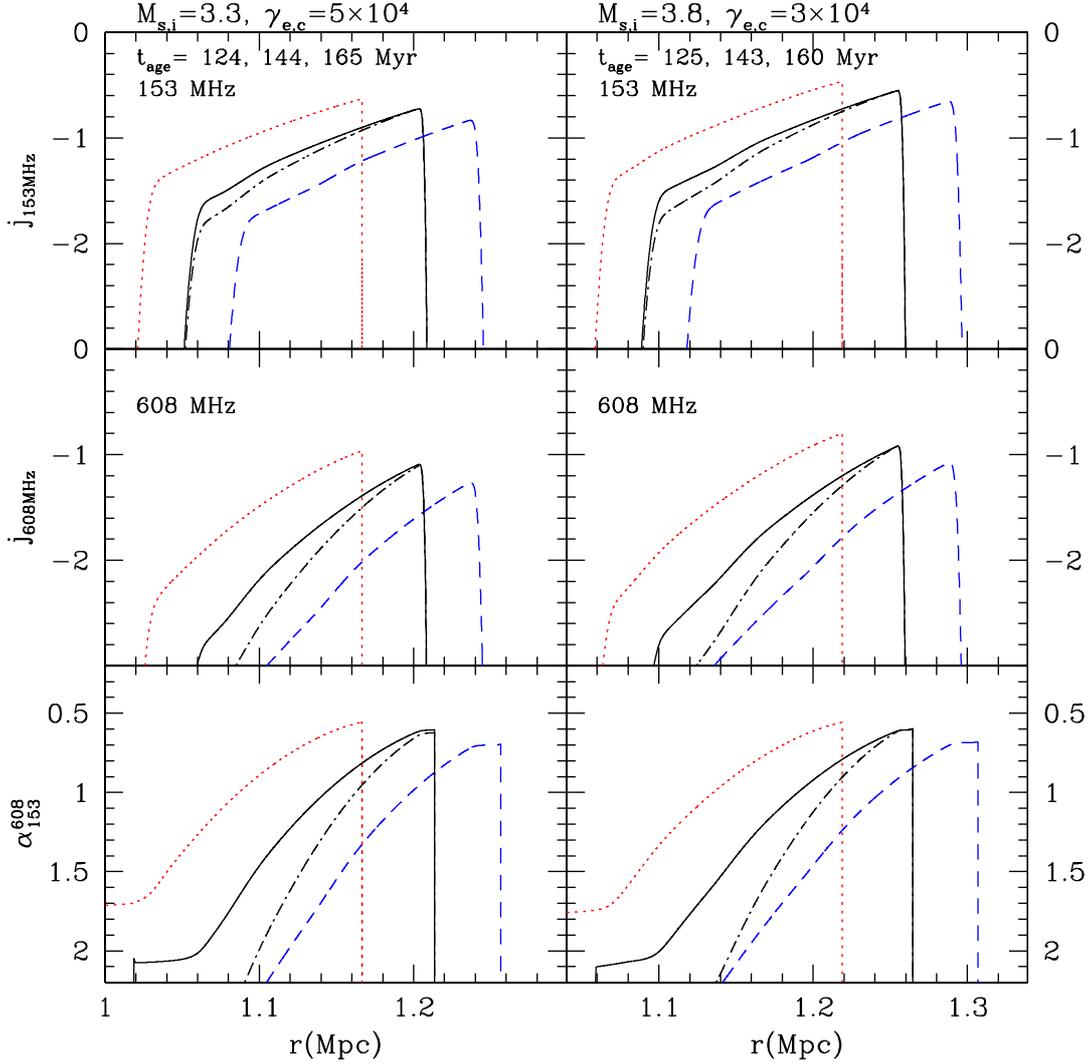}
\caption{Time evolution of the synchrotron emissivity, $j_{\nu}(r)$ at 153~MHz (top panels) and 608~MHz (middle panels), 
and the spectral index,
$\alpha_{153}^{608}$ between the two frequencies (bottom panels), plotted as a function of the radial distance from the cluster center, $r(\rm Mpc)$,
for the M3.3 model (left panels) and the M3.8 model (right panels).
The line types and the corresponding epochs are the same as in Fig. 2.
The black dot-dashed lines show the results of the same models except higher magnetic field strength, $B_1=2.5\muG$, at the
same $t_{\rm obs}$.
}
\end{figure*}

\begin{figure*}[t!]
\vskip -0.5cm
\centering
\includegraphics[trim=2mm 2mm 2mm 2mm, clip, width=150mm]{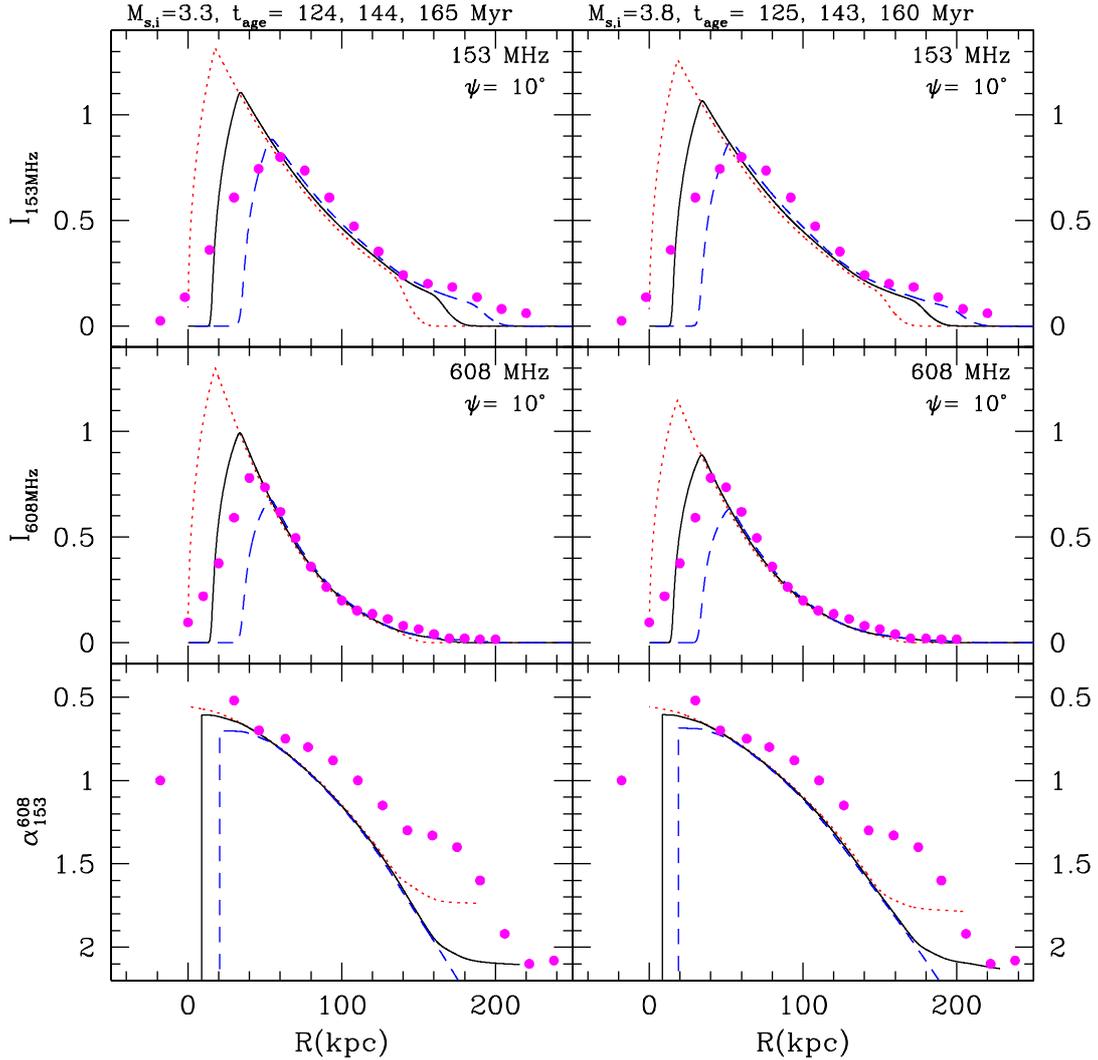}

\caption{Time evolution of the surface brightness $I_{\nu}(R)$ at 153~MHz (top panels) and at 608~MHz (middle panels),
and the spectral index $\alpha_{153}^{608}$ between the two frequencies (bottom panels) plotted 
as a function of the projected distance behind the shock, $R(\rm kpc)$
for the M3.3 model (left panels) and the M3.8 model (right panels).
The line types and the corresponding time are the same as in Fig. 2.
The projection angle, $\psi=10^{\circ}$ is adopted.
The magenta dots are the observational data taken from \citet{donnert16}.
The magenta dots are the observational data taken from \citet{donnert16}.
Note that, for the second and third epochs,
the relic edge is located behind the shock (at $R=0$).
}
\end{figure*}

\begin{figure*}[t!]
\vskip -0.5cm
\centering
\includegraphics[trim=2mm 2mm 2mm 2mm, clip, width=150mm]{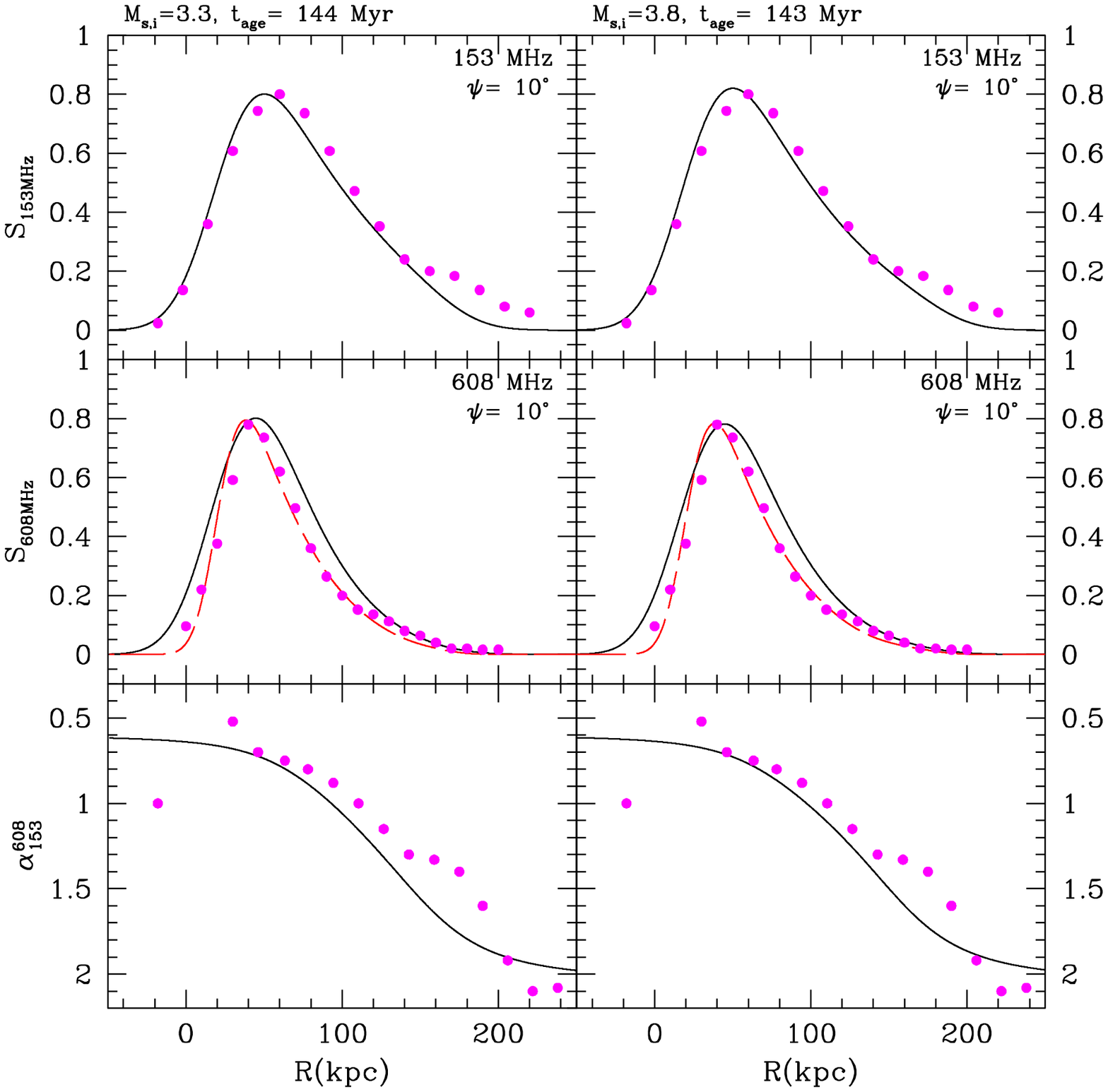}
\caption{Beam convolved brightness profile $S_{\nu}(R)$ at 153~MHz (top panels) and at 608~MHz (middle panels),
and the spectral index $\alpha_{0.15}^{0.61}$ between the two frequencies (bottom panels)
at $t_{\rm obs}=144$~Myr for the M3.3 model (left panels) 
and at $t_{\rm obs}=143$~Myr for the M3.8 model (right panels).
The simulated brightness profiles, $I_{\nu}(R)$, shown in Fig. 4 are smoothed with 
Gaussian smoothing with $50$~kpc width in order to emulate radio observations.
The brightness profile at 608~MHz smoothed with $25$~kpc width is also shown by the red long dashed lines
in the middle panels.
}
\end{figure*}

Note that the downstream flow speed in these models is smaller than the minimum value
required in the in-situ injection model considered by \citet{donnert16}.
So we consider several models with a range of parameters for the preshock magnetic field strength
and the cutoff Lorentz factor.
The preshock magnetic field strength is assumed to be $B_1= 1\muG$ in both models, 
resulting in the postshock strength, $B_2\approx 2.3-2.5\muG$,
since the factor $Q=0.65$ has the greatest value for $B_2=2.5 \muG$.
The postshock magnetic field strength is assumed to scale with the 
gas pressure, $B_{\rm dn}(r)\propto \sqrt{ P_g(r)}$, as in Paper I.

We find that the models with $B_1= 1\muG$ and $\gamma_{e,c}\approx 3-5\times 10^4$ produce the profiles of 
$S_{\nu}$ and $\alpha_{153}^{608}$ that are consistent with the observed profiles given in \citet{donnert16}.
If we take smaller values for $\gamma_{e,c}$ or larger values for $B_1$, for example, 
the simulated spectral index increases behind the shock
too fast, compared to the observed profile of $\alpha_{153}^{608}$.

If we take the mean value the observed value, $kT_1=2.7_{-0.4}^{+0.7}$~keV \citep{akamatsu15},
then the relevant velocities such as $u_s$ and $u_2$ and the downstream length scale of radio emitting
region will decrease by a factor of $\sqrt{3.4/2.7}$.
For such a model, a higher value of $\gamma_{e,c}$ would be required to increase the relic width
to the level that matches the observations.

In order to reproduce the steep spectral curvature above 2~GHz, 
we assume that the cloud with pre-existing electrons has a finite size, as in \citet{kangryu16}.
Table 1 also lists the cloud size, $L_{\rm cloud}$, and the time when the shock exits the cloud, $t_{\rm exit}$.
Considering that the observed downstream length scale is greater than 150~kpc and the shock compression
ratio is $\sigma\approx 3$, $L_{\rm cloud}\gtrsim 400$~kpc is required. 
Note that this is much greater than $L_{\rm cloud}\sim 130$~kpc adopted in \citet{kangryu16}.

We find that both the simulated brightness profiles and the integrated spectra 
become consistent with the observations at the shock age of $t_{\rm obs}\approx 144$~Myr in M3.3 model
and at $t_{\rm obs}\approx 143$~Myr in M3.8 model.
The degree of spectral steepening above $\sim2$~GHz is controlled by the time elapse between $t_{\rm exit}$
and $t_{\rm obs}$.
At $t_{\rm obs}$ the spherical shock slows down to $M_{\rm s,obs}\approx 2.7$ with $kT_{\rm 2,obs}=10.7$~keV 
in M3.3 model,
and $M_{\rm s,obs}\approx 3.1$ with $kT_{\rm 2,obs}=12.9$~keV in M3.8 model. 

As mentioned in the Introduction, \citet{kangryu16} were able to reproduce the observed $S_{\nu}$ at 610~MHz 
(with FWHM $~\sim 55$~kpc) and
the integrated spectrum of the Sausage relic with the re-acceleration models with $kT_1=3.35$~keV,
$M_{s,i}\approx3.0-3.3$, $s=4.2$, $\gamma_{e,c}= 10^4$, $B_1= 2.5\muG$, $L_{\rm cloud}\approx 131$kpc, 
and $t_{\rm obs}\approx 55$~Myr.
However, such models cannot be consistent with the brightness profiles at lower frequencies (e.g., 150~MHz) 
that extend beyond 150~kpc behind the shock.
This demonstrates that it is important to have observational data for brightness profiles 
at multi frequencies in addition to the integrated spectrum in order to constrain the best-fit DSA models.

\begin{figure*}[t!]
\vskip -0.5cm
\centering
\includegraphics[trim=2mm 3mm 2mm 2mm, clip, width=150mm]{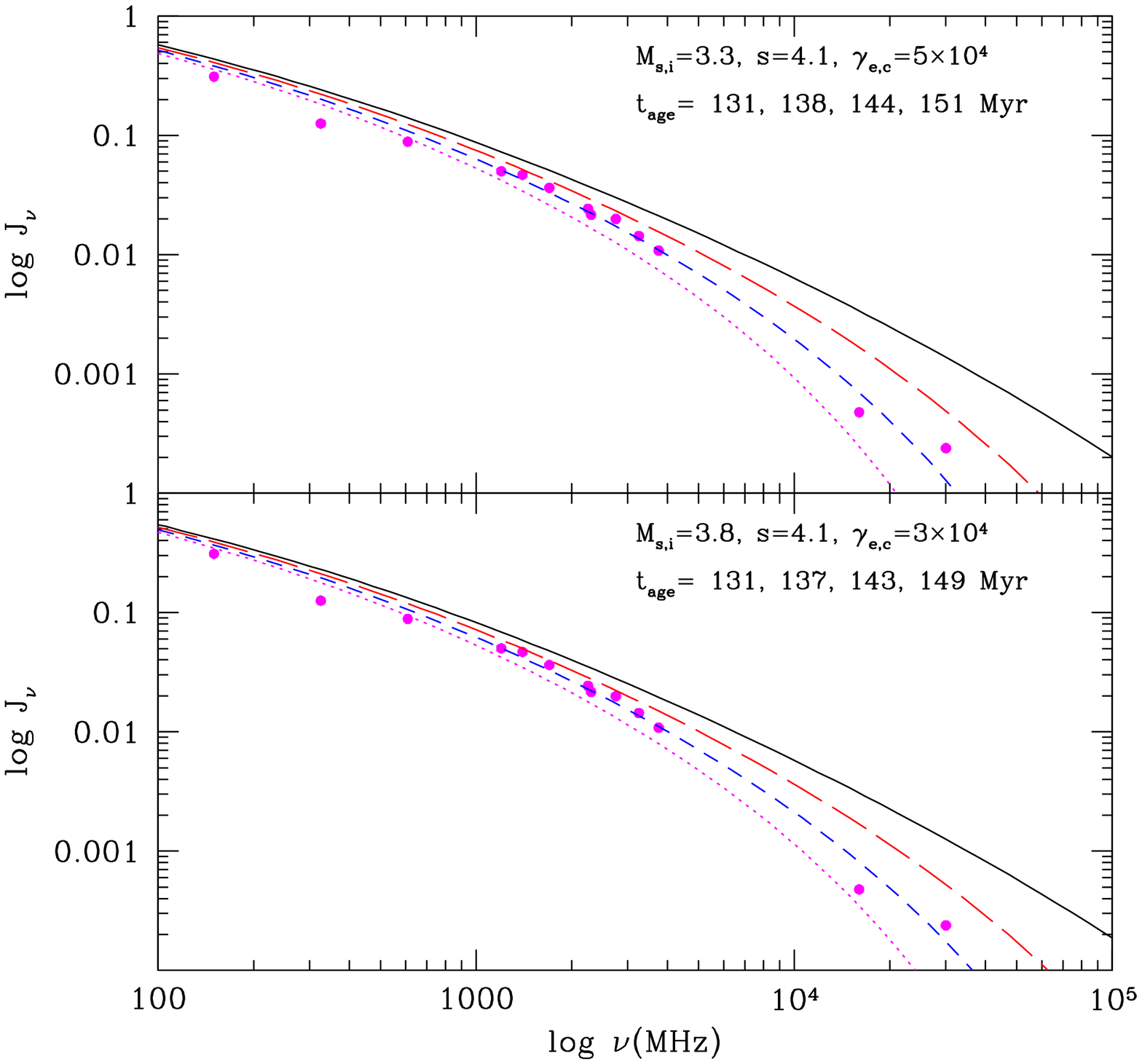}
\caption{Time evolution of volume-integrated radio spectrum 
for the M3.3 model with $M_{s,i}=3.3$ (top panel) 
at 131 (black solid line), 138 (red long-dashed), 144 (blue dashed), and 151~Myr (magenta dotted),
and 
for the M3.8 model with $M_{s,i}=3.8$ (bottom panel) 
at 131 (black solid line), 137 (red long-dashed), 143 (blue dashed), and 149~Myr (magenta dotted).
The magenta dots are the observational data taken from Table 3 of \citet{stroe16}.
}
\end{figure*}

\section{RESULTS OF DSA SIMULATIONS}

Figure 2 shows the flow speed, $u(r)$, and the magnetic field strength, $B(r)$, at three epochs:
$t_{\rm age}=124,~144,~165$~Myr for M3.3 model and $t_{\rm age}=125,~143,~160$~Myr for M3.8 model.
Here $r$ is the radial distance from the cluster center.
In both models, the shock is about to exist the cloud of pre-existing electrons at $t_{\rm exit}=124-125$~Myr
(red dotted lines), while the profiles of $\alpha_{153}^{608}$ and the integrated spectra 
become consistent with the observations of 
\citet{stroe16} at $t_{\rm obs}=143-144$~Myr (black solid lines) (see Figure 6).
Note that the downstream flow speed in the shock rest frame increases behind the shock,
while the magnetic field strength decreases in the downstream region. As a result, the cooling length,
$\Delta l_{\rm high}$, is somewhat greater than estimated for 1D planar shocks with uniform $u(r)$ and $B(r)$
in the postshock region.

Figure 3 shows the synchrotron emissivity, $j_{\nu}(r)$ at 153~MHz and 608~MHz in arbitrary units, and the spectral index,
$\alpha_{153}^{608}$ in the two models.
Since the shock exits the cloud of pre-existing electrons at $124-125$~Myr,
the edge of the radio relic is located slightly behind the shock for the second (black solid lines)
and third (blue dashed line) epochs.
For example, the relic edge is at 1.162, 1.212, and 1.242~Mpc,
while the shock position is at 1.162, 1.22, and 1.28~Mpc in M3.3 model.

From the profiles of $j_{\nu}(r)$ at 153~MHz, one can see that 
the advection length is $140-160$~kpc, consistent with Equation (\ref{lwidth1}).
Due to faster cooling of higher energy electron $j_{\nu}(r)$ at 608~MHz decreases much faster
with the FWHM about 50~kpc.
Since the shock weakens in time, the spectral index, $\alpha_{153}^{608}$, at the relic edge decreases 
from 0.55 to 0.7 during $125-165$~Myr.
In fact, the gradient of $\alpha_{153}^{608}(r)$ depends on the parameters, $s$ and $\gamma_{e,c}$
as well as $M_s(t)$ and $B_1$.
For smaller values of $\gamma_{e,c}$ or larger values of $B_1$, 
$\alpha_{153}^{608}(r)$ would increase behind the shock faster than that in the fiducial models shown in Figure 3.
For example, the black dot-dashed lines show the results of the same models at the same $t_{\rm obs}$
except higher magnetic field strength, $B_1=2.5\muG$ and $B_2\approx 5.8-6.3\muG$.

\subsection{Surface Brightness and Spectral Index Profiles}

The radio surface brightness, $I_{\nu}(R)$, is calculated from the emissivity $j_{\nu}(r)$ 
by adopting the same geometric volume of radio-emitting electrons
as in Figure 1 of \citet{kang15b}.
Here $R$ is the distance behind the projected shock edge in the plane of the sky.

Figure 4 shows $I_{\nu}(R)$ at 153~MHz and 608~MHz in arbitrary units at the same time epochs as in Figure 3.
Here the projection angle $\psi=10^{\circ}$ is adopted.
In addition, the spectral index, $\alpha_{153}^{608}$ is calculated from the projected
$I_{\nu}(R)$ at 153 and 608~MHz and shown in the bottom panels.
The magenta dots are observational data read from Figures 5 and 6 in \citet{donnert16}.
Note that the relic edge is located $20-40$~kpc behind the shock at the two later epochs (black solid and 
blue dashed lines),
since the shock breaks out of the cloud at $124-125$~Myr (red dotted lines).

The simulated brightness profiles can be compared with the observed data,
only if the vertical scale is adjusted,
since the normalization of radio brightness is arbitrary. 
Although the spatial profiles of $I_{\nu}$ at the two frequencies for $R>40$~kpc seem to match reasonably well
the observed profiles, $\alpha_{153}^{608}$ decreases somewhat faster than the observed trend.

In fact, the observed profiles should be compared with the brightness profiles convolved with telescope beam profile
with a finite width.
So we present the brightness profiles, $S_{\nu}(R)$, smoothed by Gaussian smoothing with 50~kpc width (a beam of 16 aresec)
at $t_{\rm obs}=143-144$~Myr in Figure 5.
The simulated profiles of $S_{\rm 153MHz}(R)$ seem compatible with the observed data up to 150~kpc,
beyond which they could be contaminated by the contributions from the radio galaxies (B, C, and D)
in the downstream region.
For 608~MHz, $S_{\rm 608MHz}$ smoothed with 25~kpc width seems to match much better the observed profile.
In the bottom panels, we also present $\alpha_{153}^{608}$ calculated with $S_{\nu}$ at the two frequencies,
smoothed with 50~kpc width.
The figure demonstrates that our model predictions convolved with appropriate beam widths are in reasonable 
agreement with the observations.

\subsection{Integrated Spectrum}

According to Equation (\ref{fbr}), the break frequency at $t_{\rm obs}=143-144$~Myr is $\nu_{\rm br}\approx 130$~MHz. 
So in the in-situ injection model without pre-existing electrons, the integrated radiation spectrum is expected to steepen 
from $\alpha_{\rm sh}$ to $\alpha_{\rm sh}+0.5$ gradually over $13~{\rm MHz}-1.3 {\rm GHz}$,
which is in contradiction with the observed spectrum shown in Figure 6.
In the re-acceleration model, however, the integrated spectra depends also on the spectral shape of the 
preshock electron population. In the case of the cloud of pre-existing electrons with a finite-size,
the degree of the spectral steepening also depends on the time elapse, $t_{\rm obs}-t_{\rm exit}$.
So the spectral curvature of the observed spectrum can be reproduced by adjusting the set of
model parameters, i.e., $M_s$, $B_1$, $t_{\rm exit}$, $t_{\rm obs}$, $s$, and $\gamma_{e,c}$ in our models.

Figure 6 shows how the integrated spectrum, $J_{\nu}$, steepens at higher frequencies 
due to lack of pre-existing seed electrons as well as radiative cooling during $131-151$~Myr in the two shock models.
Note again $t_{\rm exit}=124-125$~Myr.
The magenta dots show the observational data taken from Table 3 of \citet{stroe16},
which are rescaled to fit the simulated spectrum near 1~GHz by eye. 
We find that $\gamma_{e,c}=3-5\times 10^4$ is needed in order to reproduce $J_{\nu}$ both
near 1~GHz and $16-30$~GHz simultaneously at $t_{\rm obs}$.

\section{Summary}

The Sausage radio relic is unique in several aspects.
Its thin arc-like morphology and uniform surface brightness along the relic length over 2~Mpc
could be explained by the re-acceleration model in which a spherical shock sweeps through
an elongated cloud of the ICM gas with pre-existing relativistic electrons \citep{kangryu15}.
Moreover, the re-acceleration model can resolve the discrepancy between $M_{\rm radio}\approx 4.6$ 
inferred from the radio spectral index \citep{vanweeren10} and $M_{\rm X-ray}\approx 2.7$ estimated from 
X-ray temperature discontinuities \citep{akamatsu15}.
Note that in this model the spectral index at the relic edge, $\alpha_{\rm sh}\approx 0.6$,
can be controlled by the power-law index of the pre-existing electron population, $s\approx 4.1-4.2$,
independent of the shock Mach number.
The steep spectral steepening above $\sim2$~GHz \citep{stroe16} could be understood, if we assume that
the cloud of pre-existing electrons has a finite width and the shock has existed the cloud
about $10-20$~Myr ago \citep{kangryu16}.

In this study, we attempt to reproduce the observed profiles of the surface brightness at 153 and 608~MHz
and the spectral index between the two frequencies presented in \citet{donnert16}, 
using the same re-acceleration model but with a set of shock parameters different from \citet{kangryu16}.
In particular, the observational facts that $S_{\rm 153MHz}$ and $\alpha_{153}^{608}$ extend beyond 150~kpc
downstream of the shock and the degree of spectral steepening of the integrated spectrum at high frequencies provide strong
constrains to the model parameters, which are listed in Table 1.
Since the re-accelerated electron spectrum depends on the pre-existing electron population,
we find that the cutoff Lorentz factor should be fine tuned as $\gamma_{e,c}\approx 3-5\times 10^4$
in order to match the observations.

This study illustrates that it is possible to explain most of the observed properties of the Sausage relic
including the surface brightness profiles and the integrated spectrum
by the shock acceleration model with pre-existing electrons.
If the shock speed and Mach number are specified by X-ray temperature discontinuities,
the other model parameters such as magnetic field strength and the spectral shape of pre-existing electrons
can be constrained by the radio brightness profiles at multi frequencies.
Moreover, the degree of spectral steepening in the integrated spectrum at high frequencies can be modeled 
with a finite-sized cloud with pre-existing electrons.

We assume the shock has existed the cloud of pre-existing electrons at $t_{\rm exit}\approx 124-125$~Myr after
crossing the length of the cloud, $L_{\rm cloud}=367$~kpc in M3.3 model and $L_{\rm cloud}=419$~kpc in M3.8 model.
Although both M3.3 and M3.8 models produce the results comparable to the observations as shown in Figures 5 and 6,
M3.3 model seems more consistent with X-ray observations: $M_{\rm s,obs}=2.7$,
$kT_{\rm 2,obs}=10.7$~keV, and $u_{\rm s,obs}=2.6\times10^3\kms$ at the time of observation, $t_{\rm obs}\approx 144$~Myr.

However, it is not well understood how an elongated cloud of the thermal gas with such pre-existing relativistic electrons
could be generated in the ICM of CIZA J2242.8+5301.
It could be produced by strong accretion shocks or infall shocks($M_s\gtrsim 5$) in the cluster outskirts \citep{hong14} or 
by turbulence induced by merger-driven activities \citep{brunetti2014}.
Alternatively, it could originate from nearby radio galaxies, such as radio galaxy H at the eastern
edge of the relic or radio galaxies B, C, and D downstream of the relic. 
Again it is not clear how relativistic electrons contained in jets/lobes of radio galaxies 
are mixed with the background gas instead of forming a bubble of hot buoyant plasma.
Note that the shock passage through such relativistic plasma is expected to result in
a filamentary or toroidal structure \citep{ensslin02,pfrommer11}, which is inconsistent with
the thin arc-like morphology of the Sausage relic.
In conclusion, despite the success of the re-acceleration model in explaining many observed properties of the
Sausage relic, the origin of pre-existing relativistic electrons needs to be investigated further.

On the other hand, the in-situ injection model for radio relics has its own puzzles: 
(1) $M_{\rm radio}>M_{\rm X-ray}$ in some relics,
(2) low DSA efficiency expected for weak shocks with $M_s<3$,
(3) relatively low fraction of merging clusters with detected radio relics, 
compared to the theoretically expected frequency of shocks in the ICM, and
(4) some observed X-ray shocks without associated radio emission.
In particular, the generation of suparthermal electrons via wave-particle interactions, and
the ensuing enhancement of the injection to the DSA process in high beta ICM plasma
should be studied in detail by fully kinetic plasma simulations.

\acknowledgments{
This work was supported by a 2-Year Research
Grant of Pusan National University.
}


\end{document}